\documentclass[useAMS,usegraphicx]{mn2e}
\usepackage{rotate}
\usepackage{times}
\newif\ifAMStwofonts
\AMStwofontstrue
\def\gs{\mathrel{\hbox{\rlap{\hbox{\lower4pt\hbox{$\sim$}}}\hbox{$>$}}}}
\def\ls{\mathrel{\hbox{\rlap{\hbox{\lower4pt\hbox{$\sim$}}}\hbox{$<$}}}}

\def\chandra{{\it Chandra}}

\def\rosat{{\it ROSAT}}

\def\asca{{\it ASCA}}

\def\xmm{{\it XMM-Newton}}

\def\et{{et al.\ }}

\def\3c{{3C~273}}

\def\ka{{K$\alpha$}}

\def\nh{{N_{\rm H}}}

\def\cm{{\rm\thinspace cm}}
\def\erg{{\rm\thinspace erg}}
\def\eV{{\rm\thinspace eV}}

\def\ga{{\rm\thinspace gauss}}

\def\keV{{\rm\thinspace keV}}
\def\km{{\rm\thinspace km}}

\def\s{{\rm\thinspace s}}
\def\ks{{\rm\thinspace ks}}
\def\ps{{\rm\thinspace s^{-1}}}

\def\cts{{\rm\thinspace count}}

\def\cps{\hbox{$\cts\s^{-1}\,$}}

\def\ergpscmps{\hbox{$\erg\cm^{-2}\s^{-1}\,$}}

\def\ergps{\hbox{$\erg\s^{-1}\,$}}

\def\kmps{\hbox{$\km\ps\,$}}

\def\pscm{\hbox{$\cm^{-2}\,$}}

\title[\xmm\ observations of bright \rosat\ selected AGN]
      {
\xmm\ observations of bright \rosat\ selected active galactic nuclei with low 
intrinsic absorption
      }

\author[L. C. Gallo et al.]
       {L. C. Gallo,$^1$
	I. Lehmann,$^1$
	W. Pietsch,$^1$
	Th. Boller,$^1$
	W. Brinkmann,$^1$
\newauthor 
	P. Friedrich$^1$ 
	and D. Grupe$^{2}$  \\
$^{1}$ Max-Planck-Institut f\"ur extraterrestrische Physik, Postfach 1312, 85741 Garching, Germany \\
$^{2}$ Department of Astronomy and Astrophysics, The Pennsylvania State University, 525 Davey Lab, University Park, PA 16802, USA
}
\date{Accepted. Received. }
\pagerange{\pageref{firstpage}--\pageref{lastpage}}
\pubyear{2005}
\begin{document}
\maketitle
\label{firstpage}

\begin{abstract}
We present a sample of twenty-one \rosat\ bright active galactic nuclei (AGN), 
representing a range of
spectral classes, and selected for follow-up snap-shot observations 
with \xmm.  The typical exposure was between $5-10\ks$. 
The objects were primarily selected on the bases of X-ray brightness and not on
hardness ratio; thus the sample cannot be strictly defined as a ``soft''
sample.  One of the main outcomes from the \xmm\ observations was that all
of the AGN, including eleven type 1.8--2 objects,
required low levels of intrinsic absorption ($\nh \ls 10^{21}\pscm$).
The low absorption in type 2 systems is a challenge to account for
in the standard orientation-based unification model, and we discuss possible 
physical and geometrical models which could elucidate the problem.
Moreover, there does not appear to be any relation between the
strength and shape of the soft excess, and the spectral classification of the AGN in 
this sample.
We further identify a number of AGN which deserve deeper
observations or further analysis: for example, the LINERs NGC~5005 and
NGC~7331, where optically thin thermal and extended emission is detected,
and the narrow-line Seyfert 1 II~Zw~177, which shows a broad emission feature
at $\sim5.8\keV$.

\end{abstract}


% --------------------------------------------------------------------------

\section{Introduction}
\label{sect:intro}

Orientation-based unification models adopt the same physical 
emission process in all active galactic nuclei (AGN), attributing the
apparent diversity to line-of-sight obscuration.  The crux of the model
is a dense, obscuring torus.  When our line-of-sight crosses the 
torus, emission from the accretion disc and broad line region (BLR) is
obscured, and we identify objects as type 2 AGN.
When we see the torus face-on, our view of the accretion disc and BLR
is unobscured and we recognise a type 1 AGN.
There is substantial evidence, mostly based on the detection of the BLR in
polarised light (e.g. Antonucci \& Miller 1985),
that this model is correct in distinguishing between type 1 and type 2 AGN.

There are two pieces of compelling X-ray evidence in support of this model
(e.g. Awaki \et 1991; Moran \et 2001).
First, in most type 2 Seyferts the direct continuum below $2-3\keV$ is highly
absorbed, indicating that the X-rays transverse a large column density 
($\nh \gs 10^{23}\pscm$) as expected from a torus.  Secondly, a reflection
component is often detected in Seyfert 2s.  It is  
primarily identified by the presence of a strong Fe~\ka\ fluorescence
line, which is produced when the primary X-rays reflect off the cold inner walls
of the putative torus.
Heavily absorbed AGN are not likely to exhibit
many of the X-ray properties seen in type 1 AGN (e.g. relativistic lines, 
rapid variability, soft excess emission), which are associated with the
inner-most regions of the accretion disc.

Recently, X-ray observations have been presented which complicate matters.
Objects which are clearly identified by optical means as type 2 AGN;
nevertheless exhibit type 1 AGN behaviour in the X-rays
(e.g. Risaliti \et 2005; Barcons \et 2003; Matt \et 2003, Panessa \& Bassani 2002, hereafter
PB02; Boller \et 2002; Pappa \et 2001).
The apparent lack of emission from the BLR and low intrinsic absorption
measured in the X-rays is a challenge to explain with orientation-based unified 
models.  It requires that we preferentially obscure the BLR
while viewing the accretion disc directly.
%Tran (2001) suggested that there could exist a
%type 2 AGN dichotomy in which, for example, a broad line region is simply 
%absent rather than obscured (see Nicastro \et 2003)
%Lumsden \et (2004) argue against this claim

%However, there is also
%significant evidence for a type 2 AGN dichotomy in which, for example,
%a broad line region is simply absent rather than obscured (e.g.
%Nicastro \et 2003; Tran 2001).

We present a small sample of AGN observed with \xmm\ (Jansen et al. 2001)
that cover a range of spectral types. 
The objects were all detected with \rosat\ and most were optically observed
in a follow-up survey by Pietsch \et (1998) and Bischoff (2004).
The benefit of such a soft X-ray selected sample is that it allows the
identification and investigation of intrinsic low-absorption type 2 systems,
which do not clearly fit into the orientation-based unification scheme.
In addition, we identify a number of AGN, which are individually interesting
and certainly warrant deeper observations.
%As most of the sources were observed in short exposures, the main goal of
%this study is to characterise the sample and motivate deeper observations
%of interesting objects and the sample as a whole.

\section{Sample selection}

Cross correlation of the \rosat\ All Sky Survey (RASS; Voges \et 1999) 
source catalogue with
the Catalogue of Principal Galaxies (Paturel \et 1989) yielded 904 X-ray sources
with galaxy counterparts (Zimmermann \et 2001).  Follow-up optical spectroscopy
was conducted to identify many of the X-ray bright objects.
Many were identified for the first time as active galaxies, resulting in
the discovery of $> 100$ nearby
AGN (Pietsch \et 1998; Bischoff 2004).

We have compiled twenty-four \xmm\ observations of twenty-one AGN from this sample
with the intention of studying AGN unification.
Most of the objects were observed as part of the Guaranteed Time
allocated to the \xmm\ Telescope Scientist.  The objects were selected for
follow-up observations because they were:

\begin{itemize}

\item[(i)]
bright in the RASS energy band ($0.1-2\keV$) with count rates $> 0.1\cps$
(corresponding to a $0.5-2\keV$ flux $>10^{-12}\ergpscmps$ for a standard
AGN power law spectrum with $\Gamma=2$);
\item[(ii)]
nearby ($z < 0.129$; $<z> = 0.04$)
\item[(iii)]
observed through a relatively low Galactic column ($< 10^{21}\pscm$);
%\item[(iv)]
%covered a range of AGN spectral types.

\end{itemize}

As such, a reasonable broadband ($0.3-10\keV$) spectrum could be obtained with a
typical exposure of $5-10\ks$.  
%with the
%most distant object (PMN~J0623--6436) at $z=0.129$.
To acquire representation from all AGN spectral types a few objects were 
retrieved from the \xmm\ Science Archive.
In total the sample includes: 
six type 1 (two of which are Narrow-line Seyfert 1), 
three type 1.2 (five observations),
one type 1.5, 
six type 1.8 (seven observations), 
four type 1.9 (1 of which is a LINER), 
one type 2 (which is also a LINER). 
The above classification is based on optical analysis.

Note that this sample differs from the {\it soft} X-ray selected sample
presented in Grupe \et (2001) and Grupe \et (2004),
which did not yield any type 2 AGN.  The primary
difference (aside from brightness) being that the Grupe samples
included a hardness ratio cut, which favoured softer AGN.
Such a hardness criterion was not adopted in this study.

\section{Observations and data reduction}
\label{sect:data}

All of the observations were conducted with \xmm\ 
between 2000 August and 2002 December.  During these observations
the EPIC pn (Str\"uder et al. 2001) detector and
MOS (MOS1 and MOS2; Turner et al. 2001) cameras were operated in 
imaging mode.
For simplicity we do not present data acquired with the 
other instruments.
Observational details are found in Table~\ref{tab:log}.

The Observation Data Files 
were processed to produce calibrated event lists using the \xmm\
Science Analysis System ({\tt SAS v6.1.0}). Unwanted hot,
dead, or flickering pixels were removed as were events due to
electronic noise.  Event energies were corrected for charge-transfer
losses, and response matrices were generated 
for each spectrum using the {\tt SAS} tasks
{\tt ARFGEN} and {\tt RMFGEN}.
Light curves were extracted from these event lists to
search for periods of high background flaring.  Background flaring was 
problematic in several observations.  Data which were corrected for background
flaring are indicated in Table~\ref{tab:log}.
The source plus background photons were extracted from a
circular region with a radius of 35$^{\prime\prime}$, and the background was
selected from an off-source region with a radius of 50$^{\prime\prime}$
and appropriately scaled to the source
region.  Single and double events were selected for the pn and
single-quadruple events for the MOS. 
The data quality flag
was set to zero (i.e. events next to a CCD edge or bad pixel were omitted). 
%Pile-up effects were determined to be negligible.

\begin{table*}
\begin{center}
\caption{Observation Log. All of the observations were conducted 
between August 2000 and December 2002 with the
EPIC detectors in imaging mode.  Unless stated otherwise, the sources were detected 
above the background between $0.3-10.0 \keV$.
In columns (1), (2), and (3) we identify the AGN by name, right ascension, 
and declination,
respectively.  In column (4) we give the spectral classification of the AGN, and in
column (5) we state its redshift.  Column (6) and (7)  provide the observation date and 
the \xmm\ revolution number, respectively.
In column (8) the EPIC operating mode is
specified, where FF is Full Frame, eFF is extended Full Frame, and SW is Small Window.
The total amount of useful exposure (GTI) is shown in column (9).  Column (10) is left to
specify footnotes which give references to the spectral classification
of the AGN and other relevant information.
}
\begin{tabular}{cccccccccc}                
\hline
(1) & (2) & (3) & (4)  & (5) & (6) & (7) & (8) & (9) & (10) \\
Object  &  $\alpha_{2000}$ & $\delta_{2000}$ & Type  &  $z$  & Date & rev.  &  Window &  Exposure (s) &  Notes  \\
 &  & & & & year.mm.dd & & mode & pn/mos1/mos2 &  \\
\hline
HCG~4a         & 00 34 13.5 & -21 26 19 & S1.8 &  0.027  & 2002.05.29 & 452  &  FF  &  13661/21177/21394  &  $a$,1\\
ESO~113--~G~010 & 01 05 16.8 & -58 26 13 & S1.8 &  0.026  & 2001.05.03 & 256  &  FF  & 4071/6911/6911  & 1 \\
ESO~244--~G~017 & 01 20 19.6 & -44 07 43 & S1.5 &  0.024  & 2000.12.18 & 188  &  FF  &  17145/7182/7182  &  1 \\
Mrk~590        & 02 14 33.7 & -00 46 03 & S1 &  0.026  & 2002.01.01 & 378  &  SW   &  7006/--/10493 & 4,$f$\\
ESO~416--~G~002 & 02 35 13.4 & -29 36 18 & S1.9 &  0.059  & 2001.06.28 & 284  &  FF  & 4269/7182/7182  &  $a,c$,1\\
Mrk~609        & 03 25 25.2 & -06 08 30 & S1 &  0.034  & 2002.08.13 & 490  &  FF  &  6803/8861/8861  & 6\\
ESO~15--~IG~011 & 04 35 16.2 & -78 01 57 & S1.8 &  0.061  & 2000.09.29 & 148  &  FF  & 5025/7766/7765  & 1 \\
               & &     &         &        & 2001.10.31 & 347 &  FF  &  4230/7385/7385  &  $a$,1\\
MCG--01-13-025 & 04 51 41.4 & -03 48 34 & S1.2 &  0.016  & 2000.08.29 & 133  &  FF  &  906/7445/7445  &  $a$,1\\
               & &      &         &        & 2002.08.28 & 498  &  FF  &  2363/6421/6421  &  $a$,1\\
MCG--02-14-009 & 05 16 21.2 & -10 33 40 &  S1  &  0.028  & 2000.09.04 & 135  &  FF  &   4419/4804/4804  &  $a$,1\\
PMN~J0623--6436 &  06 23 07.7 & -64 36 21 & S1  &  0.129  & 2000.10.13 & 155  &  LW  &  5941/8591/8591  &  $a$,1\\ 
UGC~3973       & 07 42 32.9 & 49 48 30 & S1.2 &  0.022  & 2000.10.09 & 153  &  SW   &   1680/1861/1838  & 5  \\
               & &    &         &        & 2001.04.26 & 253  &  SW   &   3590/5737/5736  &  5 \\
Mrk~728        & 11 01 01.4 & 11 02 46 & S1.9 &  0.036  & 2002.05.23 & 449  &  FF  & 6150/8850/8851  &  7\\
CGCG~039--167  & 11 24 07.9 & 06 12 56 & S1.8 &  0.038  & 2002.12.19 & 555  &  FF  &  6387/8410/8410  &  8\\
NGC~5005       & 13 10 56.3 & 37 03 37 & L1.9&  0.003  & 2002.12.13 & 551  &  eFF  &  8796/13103/13118  &  3\\
CGCG~021--063  & 15 16 39.8 & 00 14 54  & S1.9 &  0.052 & 2002.08.02 & 485  &  eFF  & 8961/13182/13184  &  2\\
NGC~7158       & 21 56 56.5 & -11 39 32 & NLS1 &  0.028  & 2001.11.27 & 361  &  FF  &  4479/7337/7337  &  1\\
II~Zw~177      & 22 19 18.8 & 12 07 57 & NLS1 &  0.082  & 2001.06.07 & 274  &  FF  &  8650/11954/11954  &  $a,e$,2\\
ESO~602--~G~031 & 22 36 55.9 & -22 13 12 & S1.8 &  0.034 & 2001.05.25 & 267  &  FF  & 6136/10785/10785  &  $a$,1\\
NGC~7331       & 22 37 05.9 & 34 25 19 & L2   &  0.003  & 2002.12.08 &  549  &  eFF  & 7407/11549/11564  &  $d$,3\\
UGC~12138      & 22 40 17.7 & 08 03 16 & S1.8 &  0.025  & 2001.06.03 & 272  &  FF  &   3359/7280/7284  &  $a$,4\\
AM~2354--304~S & 23 57 28.4 & -30 27 39 & S1.2 &  0.031 & 2001.05.25 & 267  &  FF  &  4437/6250/6250  &  $a,b$,1\\
\hline
\label{tab:log}
\end{tabular}

\medskip
\raggedright
PROCESSING NOTES:  ($a$) Observation is corrected for background flaring. 
Source is detected above the background in the band ($b$) $0.3-6\keV$, ($c$) $0.3-7\keV$,
($d$) $0.3-8\keV$, ($e$) $0.3-9\keV$. ($f$) MOS1 was operated in timing mode and the data are ignored here.
REFERENCES: (1) Pietsch \et (1998). (2) Bischoff (2004). (3) Ho \et (1997). 
(4) Osterbrock \& Martel (1993). (5) Osterbrock \& Shuder (1982).
(6) Kewley \et (2001).  (7) Kuraszkiewicz \et (2004). (8) Veron-Cetty \& Veron (2003).

\end{center}
\end{table*}

% --------------------------------------------------------------------------
\section{Spectral analysis}
\label{sect:fit}

Each AGN spectrum was compared to the
respective background spectrum to determine the energy range in which the
source was reasonably detected above the background.  
Most of the objects were detected between $0.3-10\keV$.  Sources which were
not detected in this entire energy band are marked in Table~\ref{tab:log}.

The source spectra were grouped such that each bin contained at least 20
counts. Spectral fitting was performed using {\tt XSPEC v11.3.1} (Arnaud
1996). 
Fit parameters are reported in the rest frame of the object, although most
of the figures remain in the observed frame.
The quoted errors on the model parameters correspond to a 90\% confidence
level for one interesting parameter (i.e. a $\Delta\chi^2$ = 2.7 criterion).
The fit parameters, uncertainties, and statistics are reported for the
combined fits to all the EPIC data; however the figures display only
the pn data for clarity.
K-corrected luminosities were derived assuming isotropic emission.
A value for the Hubble constant of $H_0$=$\rm 70\ km\ s^{-1}\ Mpc^{-1}$ and
a standard cosmology with $\Omega_{M}$ = 0.3 and $\Omega_\Lambda$ = 0.7
were adopted.
The value for the Galactic column density toward each object was taken from
Dickey \& Lockman (1990) and is given in Table~\ref{tab:cont}.

\subsection{The broadband X-ray continuum}
\label{sect:bbc}

As a first approximation, the $0.3-10\keV$ spectrum of each object was 
fitted with a single power law which was
shaped by a fixed Galactic column density (Dickey \& Lockman 1990)
and a free intrinsic column density.  
This simple fit is plotted in Figure~\ref{fig:po} for each AGN,
and the quality of the fits are reported in column~2 
of Table~\ref{tab:cont}.  For four objects: Mrk~728, Mrk~609, CGCG~021--063,
and MCG--01-13-025, 
this simple model provided 
a statistically acceptable fit ($\chi^{2}_{\nu} \approx 1$).

To examine a multi-component broadband
continuum, the single power law was replaced by: (1) a broken power law model,
and (2) a blackbody plus power law model.  
Both multi-component continuum models provided improvement
over the single power law continuum, and in most cases, an
acceptable fit as well.  
However, for the two LINERs, NGC~5005 and NGC~7331, strong residuals
remained at low energies indicating that neither multi-component continuum
was completely satisfactory.
A significant improvement was established when
a power law plus optically thin
emission (i.e. a Raymond-Smith plasma; Raymond \& Smith 1977), 
rather than optically thick blackbody emission, was considered.     
Comparable fits were obtained when the {\tt XSPEC} model
{\tt MEKAL} 
(Mewe, Gronenschild \& van den Oord 1985;
Mewe, Lemen, \& van den Oord 1986;
Kaastra 1992)
was employed instead of the Raymond-Smith model. 

The improvement over the
single power law model obtained with the broken power law and the power law plus
blackbody (or Raymond-Smith) models are shown in column 3
and 4 of Table~\ref{tab:cont},
respectively.  

For simplicity, the thermal
continuum models are adopted for the remaining analysis in order to
characterise the individual spectra and generalise the sample.
Implications of this decision are discussed in Sect.~\ref{sect:bbc}.
The resulting best-fit parameters are provided in Table~\ref{tab:cont}.

\begin{table*}
\begin{center}
\caption{Broadband continuum fits. Continuum models are presented for each object
in the $0.3-10\keV$ band unless specified otherwise in Table~\ref{tab:log}.
The object name is given in column (1).  In column (2) the fit quality ($\chi^2_\nu$)
is stated for an absorbed (Galactic and intrinsic) power law fitted over the broadband
(fit is shown in Figure~\ref{fig:po}).  The improvement over the power law fit 
provided by a broken power law model ($\Delta\chi^2_{bk}$ for two additional
parameters: $\Gamma$ and break energy $E_b$) and a blackbody plus power law model
($\Delta\chi^2_{bb}$ for two additional parameters: blackbody temperature $kT$ and 
normalisation) is given in column (3) and (4), respectively.
Columns (5)-(11) are fit parameters and quantities which are measured by adopting the
blackbody plus power law model given in column (4).  Column (5) shows the fixed
Galactic column density ($10^{20}\pscm$).  
Column (6) shows the measured column density at the redshift
of the AGN ($10^{20}\pscm$).  
In column (7) the blackbody temperature (\eV) is given and in column (8) the
power law photon index.  
The observed $0.3-10\keV$ flux corrected for Galactic absorption
in units of $10^{-12}\ergpscmps$
is reported in column (9).  
The rest frame $2-10\keV$ luminosity in units of $10^{43}\ergps$ is estimated in
column (10).  
In column (11) the luminosity ratio between the thermal and power law component
(corrected for Galactic and intrinsic
absorption) in the $0.3-10\keV$ band is estimated.
Only the pn fluxes and luminosities are reported in columns 9--11 as there
are known cross-calibration uncertainties between MOS and pn in the model
normalisations (Kirsch 2005).
}
\begin{tabular}{ccccccccccc}                
\hline
(1) & (2) & (3) & (4)  & (5) & (6) & (7) & (8) & (9) & (10) & (11)\\
 Object  &  $\chi^{2}_{\nu}$/dof &  $\Delta\chi^2_{bk}$  &  $\Delta\chi^2_{bb}$ & $\nh_{G}$ & $\nh_{z}$  &  $kT$ &  $\Gamma$  &  $F_{0.3-10}$ & $L_{2-10}$ & $\frac{L_{th}}{L_{po}}$   \\
\hline
HCG~4a & $1.24/234$ & $51.4$ & $72.8$ & $1.55 $ & $<2.98$ & $199\pm10$ & $1.44^{+0.13}_{-0.08}$ & $0.69$ & $0.06$ & $0.24$ \\
ESO~113-~G~010 & $3.30/477$ & $1108.2$ & $1119.1$ & $2.77$ & $<0.72$ & $98\pm5$ & $1.91^{+0.04}_{-0.03}$ & $6.24$ & $0.38$ & $0.27$ \\
ESO~244-~G~017 & $1.15/872$ & $160.9$ & $114.2$ & $2.25$ & $<0.21$ & $124^{+7}_{-9}$ & $1.89^{+0.03}_{-0.05}$ & $6.08$ & $0.37$ & $0.09$ \\
Mrk~590 & $1.11/652$ & $63.9$ & $71.8$ & $2.68$ & $<0.26$ & $161\pm10$ & $1.66\pm0.03$ & $8.43$ & $0.72$ & $0.08$ \\
ESO~416-~G~002 & $1.02/634$ & $39.6$ & $39.7$ & $1.78$ & $<2.70$ & $105^{+6}_{-4}$ & $1.63^{+0.06}_{-0.04}$ & $7.53$ & $3.76$ & $0.02$ \\
Mrk~609 & $0.90/309$ & $24.6$ & $23.9$ & $4.41$ & $<0.64$ & $174\pm20$ & $1.64^{+0.03}_{-0.04}$ & $2.43$ & $0.36$ & $0.10$ \\
ESO~15-~IG~011(1) & $1.06/310$ & $17.2$ & $16.9$ & $7.57$ & $<3.84$ & $108^{+27}_{-16}$ & $1.86^{+0.03}_{-0.06}$ & $3.24$ & $1.42$ & $0.06$ \\
ESO~15-~IG~011(2) & $1.29/169$ & $24.9$ & $28.9$ & $7.57$ & $<4.07$ & $144^{+18}_{-22}$ & $1.68\pm0.10$ & $1.81$ & $0.84$ & $0.13$ \\
MCG--01-13-025(1) & $0.88/526$ & $8.8$ & $8.8$ & $4.17$ & $<1.79$ & $156\pm85$ & $1.79\pm0.05$ & $13.90$ & $0.43$ & $0.02$ \\
MCG--01-13-025(2) & $1.01/875$ & $64.7$ & $64.5$ & $4.17$ & $<0.39$ & $139^{+27}_{-32}$ & $1.70\pm0.03$ & $21.31$ & $0.70$ & $0.03$ \\
MCG--02-14-009 & $1.22/560$ &  $146.0$ & $179.3$ & $9.29$ & $4.81^{+1.06}_{-3.02}$ & $116\pm10$ & $1.85\pm0.06$ & $7.79$ & $0.66$ & $0.39$ \\
PMN~J0623-6436 & $1.26/701$ & $190.4$ & $166.8$ & $5.41$ & $<0.27$ & $108^{+7}_{-10}$ & $1.96\pm0.04$ & $8.43$ & $15.55$ & $0.15$ \\
UGC~3973(1) & $1.27/774$ & $229.6$ & $252.2$ & $5.68$ & $<1.65$ & $112\pm7$ & $1.82\pm0.03$ & $43.64$ & $2.27$ & $0.13$ \\
UGC~3973(2) & $2.46/924$ & $1256.2$ & $1269.6$ & $5.68$ & $<3.42$ & $100\pm5$ & $1.67\pm0.03$ & $21.66$ & $1.30$ & $0.29$ \\
Mrk~728 & $0.98/687$ & $18.6$ & $20.7$ & $2.18$ & $<0.51$ & $198\pm20$ & $1.68^{+0.03}_{-0.05}$ & $6.76$ & $1.15$ & $0.05$ \\
CGCG~039-167 & $1.06/513$ & $28.0$ & $7.9$ & $4.29$ & $<0.88$ & $157^{+48}_{-80}$ & $1.79\pm0.05$ & $4.34$ & $0.76$ & $0.04$ \\
NGC~5005 & $3.25/182$ & $218.6$ & $395.3^r$ & $1.08$ & $<1.49$ & $769^{+28}_{-32}$ & $1.58\pm0.07$ & $0.75$ & $0.0007$ & $0.29$ \\
CGCG~021-063 & $0.93/458$ & $4.8$ & $1.8$ & $4.29$ & $<1.77$ & $200^{+70}_{-66}$ & $1.69^{+0.05}_{-0.03}$ & $3.08$ & $1.13$ & $0.02$ \\
NGC~7158 & $1.13/57$ & $13.5$ & $11.7$ & $3.93$ & $<4.74$ & $109^{+19}_{-24}$ & $1.90\pm0.21$ & $0.47$ & $0.16$ & $0.25$ \\
II~Zw~177 & $1.27/542$ & $107.6$ & $146.7$ & $5.41$ & $<0.26$ & $135\pm5$ & $2.64\pm0.04$ & $5.79$ & $1.65$ & $0.27$ \\
ESO~602-~G~031 & $1.79/798$ & $595.2$ & $595.8$ & $2.15$ & $1.84^{+1.80}_{-1.18}$ & $83\pm3$ & $1.77\pm0.04$ & $9.59$ & $1.26$ & $0.13$ \\
NGC~7331 & $2.20/125$ & $65.3$ & $146.4^r$ & $8.61$ & $<5.95$ & $497^{+125}_{-56}$ & $1.79^{+0.11}_{-0.14}$ & $0.59$ & $0.0005$ & $0.22$ \\
UGC~12138 & $1.30/825$ & $265.7$ & $288.6$ & $6.68$ & $<0.82$ & $105\pm6$ & $1.87\pm0.04$ & $17.26$ & $1.13$ & $0.14$ \\
AM~2354-304 & $1.24/843$ &  $168.8$ & $175.6$ & $1.35$ & $<0.25$ & $98\pm5$ & $2.14^{+0.03}_{-0.04}$ & $9.01$ & $0.70$ & $0.15$ \\
\hline
\label{tab:cont}
\end{tabular}

\medskip
\raggedright
$r$: The temperatures given for NGC~5005 and NGC~7331 are for an optically thin
thermal plasma rather then for blackbody emission (see text for
details).

\end{center}
\end{table*}

% --------------------------------------------------------------------------

\begin{figure*}
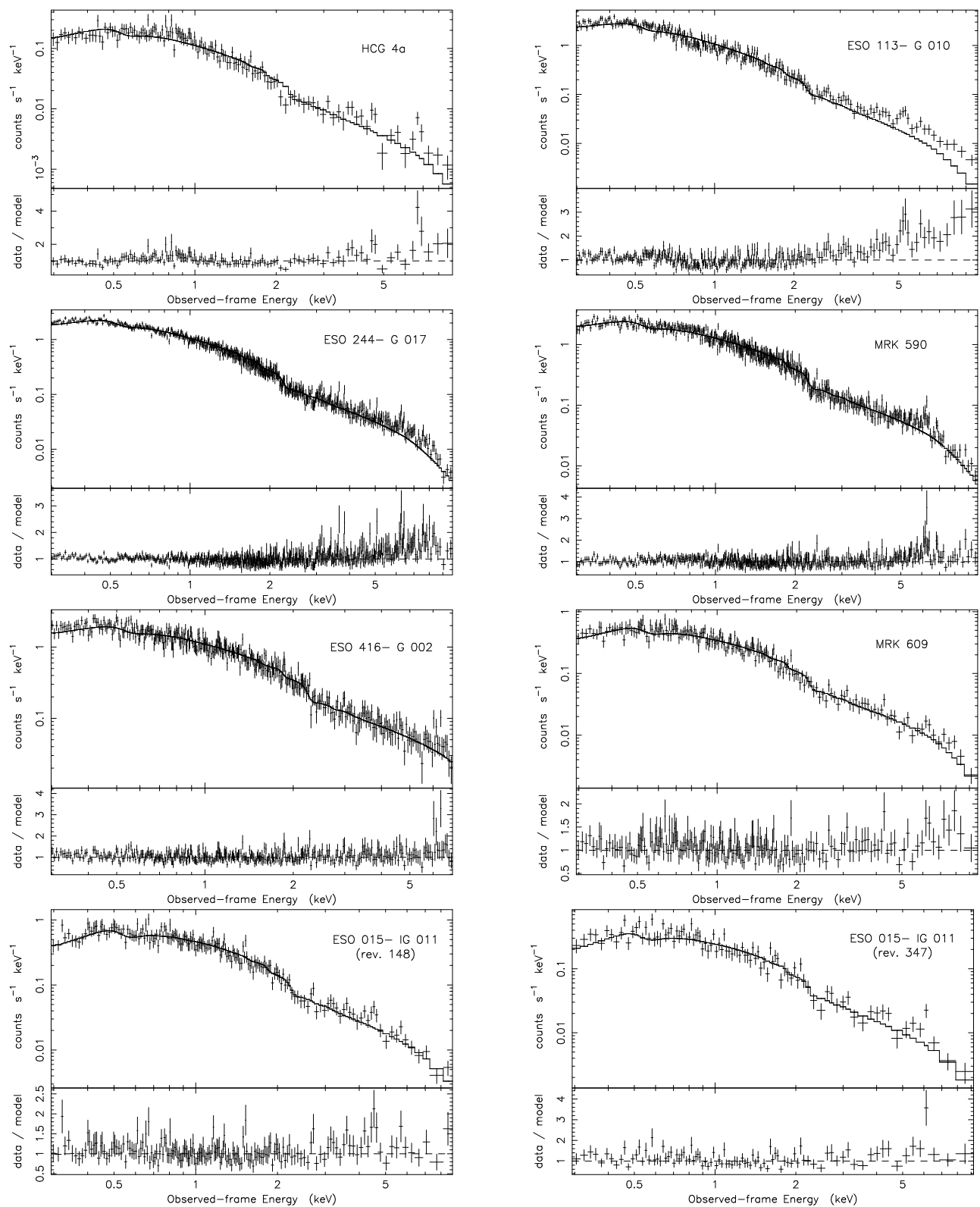

\begin{minipage}[]{0.48\hsize}
\scalebox{0.33}{\includegraphics[angle=270]{fig1a.ps}}
\end{minipage}
\hfill
\begin{minipage}[]{0.48\hsize}
\scalebox{0.33}{\includegraphics[angle=270]{fig1b.ps}}
\end{minipage}
\hfill
\begin{minipage}[]{0.48\hsize}
\scalebox{0.33}{\includegraphics[angle=270]{fig1c.ps}}
\end{minipage}
\hfill
\begin{minipage}[]{0.48\hsize}
\scalebox{0.33}{\includegraphics[angle=270]{fig1d.ps}}
\end{minipage}
\hfill
\begin{minipage}[]{0.48\hsize}
\scalebox{0.33}{\includegraphics[angle=270]{fig1e.ps}}
\end{minipage}
\hfill
\begin{minipage}[]{0.48\hsize}
\scalebox{0.33}{\includegraphics[angle=270]{fig1f.ps}}
\end{minipage}
\hfill
\begin{minipage}[]{0.48\hsize}
\scalebox{0.33}{\includegraphics[angle=270]{fig1g.ps}}
\end{minipage}
\hfill
\begin{minipage}[]{0.48\hsize}
\scalebox{0.33}{\includegraphics[angle=270]{fig1h.ps}}
\end{minipage}
\hfill
\caption
{\label{fig:po}
The broadband ($0.3-10\keV$, unless stated differently in Table~\ref{tab:log})
spectrum fitted with an absorbed (Galactic and intrinsic) power law and the
residuals (data/model) resulting from the fit.  For clarity, only the pn
data are shown.
As the fit statistics are dominated by the low-energy photons ($E \ls 3\keV$),
excess residuals will most likely appear at higher energies.
}
\end{figure*}

\setcounter{figure}{0}

\begin{figure*}
\begin{minipage}[]{0.48\hsize}
\scalebox{0.33}{\includegraphics[angle=270]{fig1i.ps}}
\end{minipage}
\hfill
\begin{minipage}[]{0.48\hsize}
\scalebox{0.33}{\includegraphics[angle=270]{fig1j.ps}}
\end{minipage}
\hfill
\begin{minipage}[]{0.48\hsize}
\scalebox{0.33}{\includegraphics[angle=270]{fig1k.ps}}
\end{minipage}
\hfill
\begin{minipage}[]{0.48\hsize}
\scalebox{0.33}{\includegraphics[angle=270]{fig1l.ps}}
\end{minipage}
\hfill
\begin{minipage}[]{0.48\hsize}
\scalebox{0.33}{\includegraphics[angle=270]{fig1m.ps}}
\end{minipage}
\hfill
\begin{minipage}[]{0.48\hsize}
\scalebox{0.33}{\includegraphics[angle=270]{fig1n.ps}}
\end{minipage}
\hfill
\begin{minipage}[]{0.48\hsize}
\scalebox{0.33}{\includegraphics[angle=270]{fig1o.ps}}
\end{minipage}
\hfill
\begin{minipage}[]{0.48\hsize}
\scalebox{0.33}{\includegraphics[angle=270]{fig1p.ps}}
\end{minipage}
\hfill
\caption
{\label{fig:po}
continued}
\end{figure*}

\setcounter{figure}{0}

\begin{figure*}
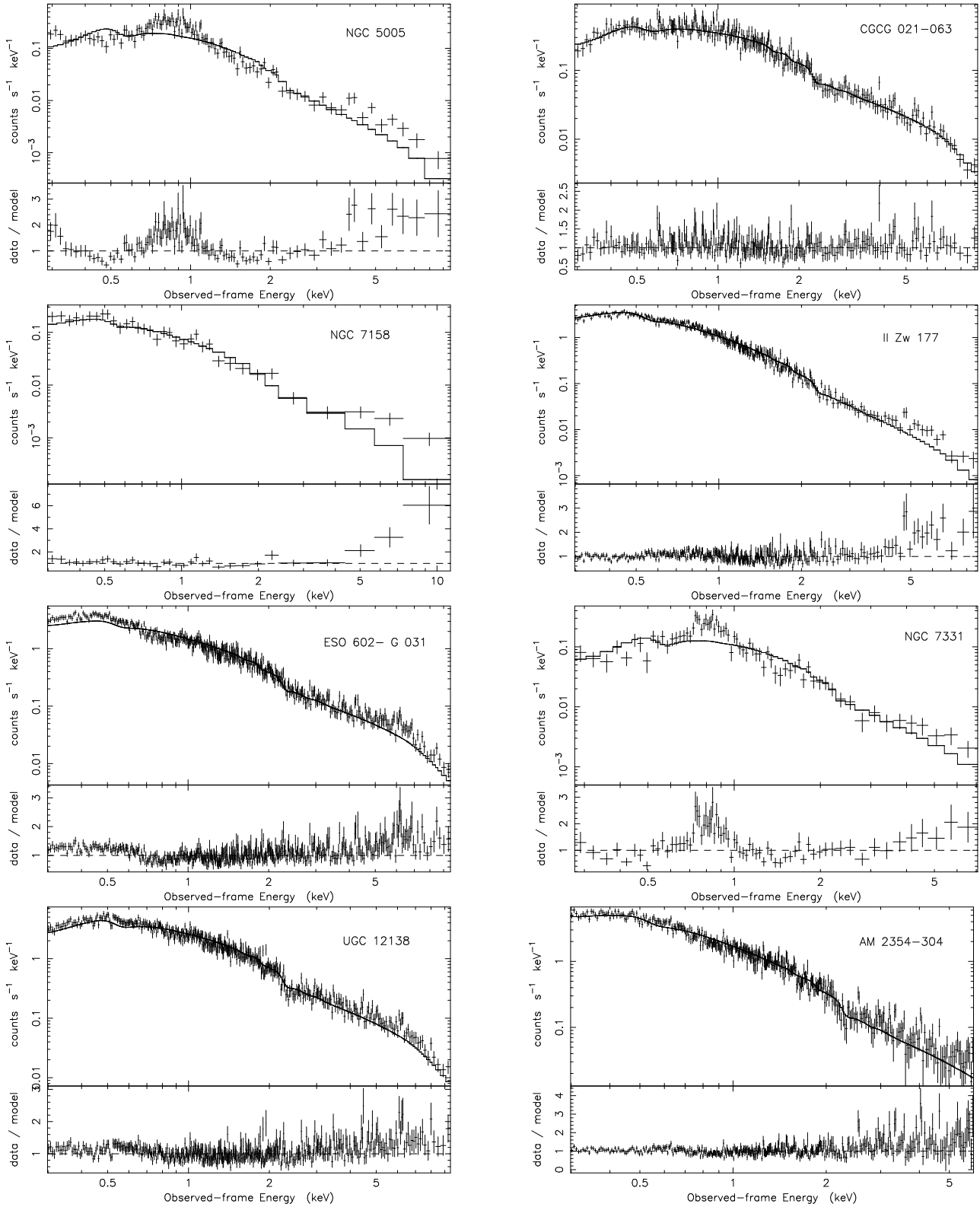

\begin{minipage}[]{0.48\hsize}
\scalebox{0.33}{\includegraphics[angle=270]{fig1q.ps}}
\end{minipage}
\hfill
\begin{minipage}[]{0.48\hsize}
\scalebox{0.33}{\includegraphics[angle=270]{fig1r.ps}}
\end{minipage}
\hfill
\begin{minipage}[]{0.48\hsize}
\scalebox{0.33}{\includegraphics[angle=270]{fig1s.ps}}
\end{minipage}
\hfill
\begin{minipage}[]{0.48\hsize}
\scalebox{0.33}{\includegraphics[angle=270]{fig1t.ps}}
\end{minipage}
\hfill
\begin{minipage}[]{0.48\hsize}
\scalebox{0.33}{\includegraphics[angle=270]{fig1u.ps}}
\end{minipage}
\hfill
\begin{minipage}[]{0.48\hsize}
\scalebox{0.33}{\includegraphics[angle=270]{fig1v.ps}}
\end{minipage}
\hfill
\begin{minipage}[]{0.48\hsize}
\scalebox{0.33}{\includegraphics[angle=270]{fig1w.ps}}
\end{minipage}
\hfill
\begin{minipage}[]{0.48\hsize}
\scalebox{0.33}{\includegraphics[angle=270]{fig1x.ps}}
\end{minipage}
\hfill
\caption
{\label{fig:po}
continued}
\end{figure*}

% --------------------------------------------------------------------------

\subsection{Additional spectral features}
\label{sect:asf}

\subsubsection{High-energy emission features}
\label{sect:hef}

The most common emission features in the $2-10\keV$ band of AGN spectra are those
of iron between $6.4-6.97\keV$ (depending on the ionisation state of
Fe).  The width and profile of the line can potentially reveal its physical origin. 
If the line is emitted from the BLR or the molecular torus
the width would be narrow and unresolved in the \xmm\ pn spectrum,
which has a resolution of $\sim 150\eV$ ($FWHM$ at 7\keV; Ehle \et 2004).
On the other hand, a significantly broadened and asymmetric profile can be manifested
from a line emitted in the inner accretion disc close to the putative supermassive
black hole.  

\begin{figure}
\rotatebox{270}
{\scalebox{0.32}{\includegraphics{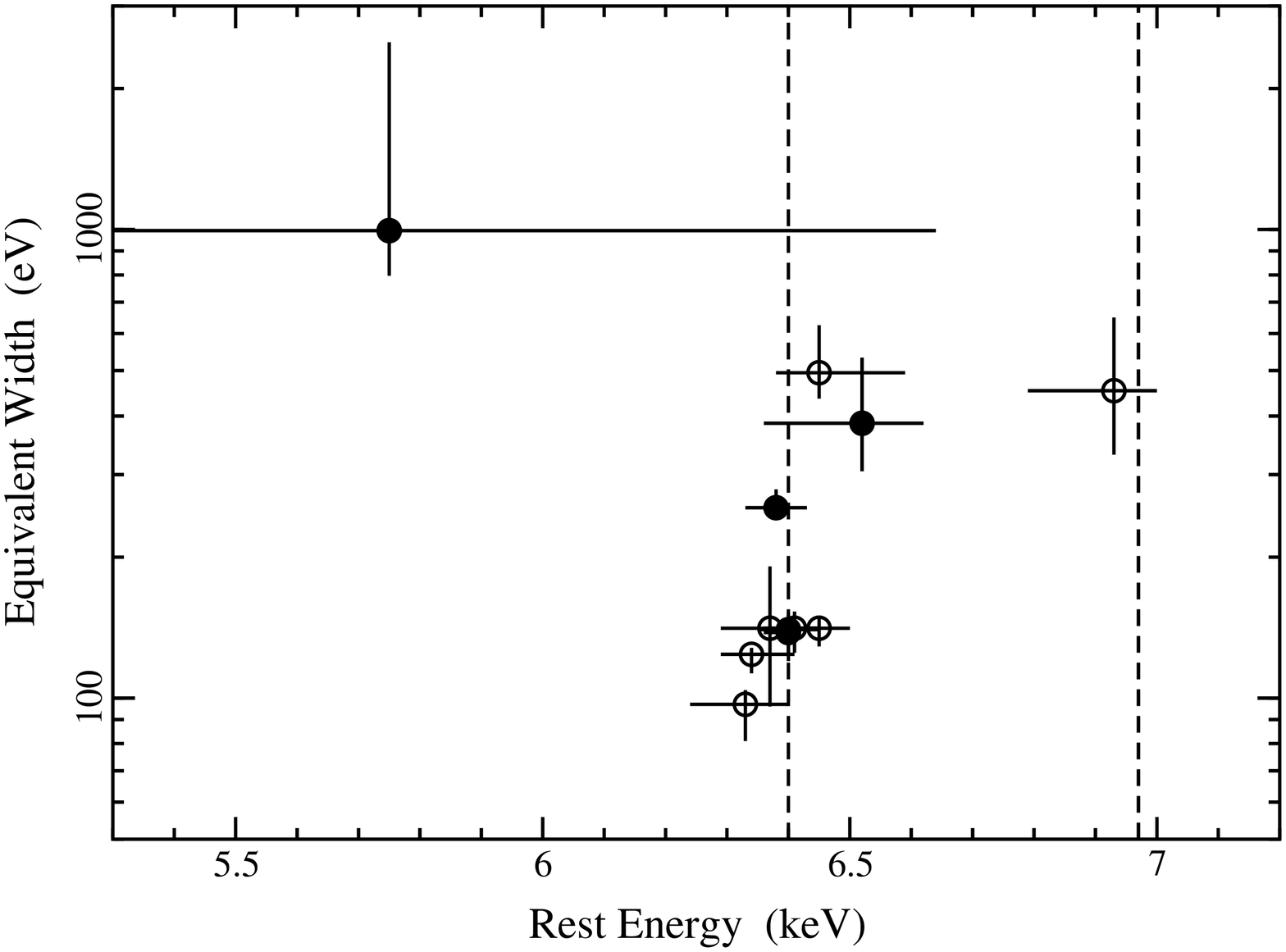}}}
\caption{The equivalent widths and rest frame energies of the detected emission
features in the sample.  
The vertical dashed lines correspond to the energy of neutral ($6.4\keV$) and
completely ($6.97\keV$) ionised Fe~\ka.  The open circles are associated with
intrinsically narrow lines and the filled circles with broad lines.  
%Error bars are
%plotted for all the data, but in some cases they are smaller than the 
%points itself.
}
\label{fig:line}
\end{figure}

Emission features in the high-energy spectra are examined for by adding
a Gaussian profile to the continuum model discussed above.  Initially
the line width was fixed intrinsically narrow ($\sigma=1\eV$).  Subsequently 
the width parameter was allowed to vary to determine if it improved the fit.
In the interest of brevity we do not test the significance of possible
detections in a robust manner (e.g. Protassov \et 2002), instead we 
claim a detection ``(marginally) significant'' if it meets the criterion
$\Delta\chi^2 > 3$ for one additional free parameter.  Further
detailed analysis of these spectra should then include a more rigorous
investigation of the significance of these line detections, in particularly
for the marginal cases.  
The objects which satisfy our criteria for a line detection are reported in
Table~\ref{tab:adds}.  
The distribution of line energies and equivalent widths in the sample
is presented in Figure~\ref{fig:line}.

% --------------------------------------------------------------------------

% --------------------------------------------------------------------------

\begin{table*}
\begin{center}
\caption{Additional spectral features on the blackbody plus power law 
continuum shown in Table~\ref{tab:cont}.  
The object name is given in column (1).  In columns (2), (3), and (4) the parameters
energy, width, and equivalent width are given for a modelled emission line (Gaussian
profile).  The improvement to the continuum fit by adding an emission line
 is given in column (5).
Columns (6) and (7) are the energy and optical depth of an absorption edge added
to the continuum model.  In column (8) the fit improvement obtained by adding an
edge is stated.
Values marked with an $f$ indicate that the parameter was fixed.
In column (5) and (8), the values in brackets indicate the number of 
free parameters added to improve the fit.
}
\begin{tabular}{cccccccc}                
\hline
(1) & (2) & (3) & (4)  & (5) & (6) & (7) & (8) \\
 Object  &  $E_{line}$ & $\sigma$ & $EW$ & $\Delta\chi^2_{line}$ 
& $E_{edge}$ & $\tau$ & $\Delta\chi^2_{edge}$ \\ 
 &  (\keV)  & (\eV) & (\eV) &  
& (\eV) &  & \\ 
\hline
HCG~4a & $6.93^{+0.07}_{-0.14}$ & $1^f$ & $453^{+196}_{-122}$ & $9.0(2)$ & -- & -- & -- \\
ESO~244-~G~017 & $6.41^{+0.04}_{-0.05}$ & $1^f$ & $141^{+12}_{-16}$ & $14.2(2)$ & -- & -- & -- \\
Mrk~590 & $6.38\pm0.05$ & $94^{+72}_{-77}$ & $255^{+24}_{-13}$ & $27.2(3)$ & -- & -- & -- \\
ESO~015-~IG~011(2) & $6.45^{+0.14}_{-0.07}$ & $1^f$ & $495^{+130}_{-59}$ & $11.2(2)$ & -- & -- & -- \\
MCG--02-14-009 & $6.52^{+0.10}_{-0.16}$ & $177^{+193}_{-118}$ & $386^{+147}_{-81}$ & $18.0(3)$ & $751\pm37$ & $0.32^{+0.14}_{-0.13}$ & $14.5(2)$ \\
UGC~3973(1) & $6.34^{+0.07}_{-0.10}$ & $1^f$ & $97^{+7}_{-16}$ & $8.5(2)$ & -- & -- & -- \\
UGC~3973(2) & $6.41^{+0.03}_{-0.04}$ & $111^{+78}_{-61}$ & $176^{+14}_{-18}$ & $23.5(3)$ & $725\pm13$ & $0.45^{+0.10}_{-0.08}$ & $51.9(2)$ \\
Mrk~728 & $6.37\pm0.08$ & $1^f$ & $141^{+50}_{-45}$ & $12.4(2)$ & -- & -- & --  \\
CGCG~039-167 & $6.45^{+0.05}_{-0.07}$ & $1^f$ & $141^{+8}_{-12}$ & $6.9(2)$ & -- & -- & -- \\
II~Zw~177 & $5.75^{+0.89}_{-0.93}$ & $1168^{+1313}_{-443}$ & $995^{+1515}_{-198}$ & $10.8(3)$ & -- & -- & -- \\
ESO~602-~G~031 & $6.40\pm0.05$ & $1^f$ & $140\pm11$ & $16.5(2)$ & $687^{+15}_{-16}$ & $0.39^{+0.13}_{-0.08}$ & $52.6(2)$  \\
UGC~12139 & $6.34^{+0.07}_{-0.05}$ & $1^f$ & $124^{+4}_{-11}$ & $14.2(2)$ & $731\pm16$ & $0.32\pm0.10$ & $30.4(2)$  \\
\hline
\label{tab:adds}
\end{tabular}
\end{center}
\end{table*}

\subsubsection{Warm absorber type features}

Imprinted on the X-ray continuum of some AGN are features
associated with absorption and/or emission from optically thin ionised
gas along the line-of-sight, the so-called warm absorber.  
It was mentioned that the two LINERs in the sample
(NGC~5005 and NGC~7331) were better fitted when emission associated
with an optically thin ionised gas was considered.  
In type 1 AGN the most prominent features associated with this
warm absorber are the O~\textsc{vii} and O~\textsc{viii} 
absorption edges at $739\eV$ and $871\eV$, respectively.

To examine warm absorption, an absorption edge was added to
the best-fit continuum model of each object.  Following the criteria of
``significance'' established for the high-energy emission 
lines (see Sect.~\ref{sect:hef}), significant detections of 
absorption edges are reported in 
four objects (Table~\ref{tab:adds}).

\section{Timing behaviour}
\label{sect:time}

An investigation of X-ray variability in the sample was not entirely practical
primarily due to the short exposures and gaps (arising from background flaring) 
in the light curves of many sources
Moreover, rapid variability is not normally expected from type 2 AGN, given the high
level of absorption usually associated with them.  However, as a matter of completeness
we did conduct a simple test for any extreme, possibly atypical, behaviour.
Light curves were created for each AGN in the $0.2-10\keV$ range and binned $100\s$.
Each light curve was then compared to a constant using a $\chi^2$-test. 
Only two light curves were found inconsistent with a constant at $> 99.9\%$.

ESO~113--~G~010, a Seyfert 1.8, demonstrated fluctuations of about $\pm15\%$ during 
the short ($< 5\ks$) observation.   In combination with the extreme soft excess and
$\sim5.4\keV$ emission feature (see Section~\ref{sect:unf}), 
the variability in this Seyfert 1.8 certainly adds
to its intrigue.

The second object was the Seyfert 1.5, ESO~244--~G~017, which had the
longest observation in our sample ($\sim 20\ks$).  
Given the duration of the observation the variability cannot be considered atypical.
However a modest flare-like event in which the flux increases by $\sim 50\%$ in 
$\sim 5000\s$ was detected, and was accompanied by mild spectral softening.
%\begin{figure}
%\rotatebox{270}
%{\scalebox{0.32}{\includegraphics{244lc.ps}}}
%\caption{Top panel: The $0.2-10\keV$ light curve of the Seyfert 1.5 
%ESO~244-~G~017 binned in $1000\s$ for illustrative purposes.
%Lower panel: The hardness ratio defined as $(H-S)/(H+S)$ as a function
%of time.  In this case $H$ is the $2.0-10\keV$ count rate, and $S$ is the
%$0.2-2.0\keV$ count rate.  The average hardness ratio is marked by the 
%dashed line.  The time axis marks elapsed time from the start of the observation.
%}
%\label{fig:244lc}
%\end{figure}

\section{Discussion}
\label{sect:diss}

%\section{Sample properties}
%\label{sect:prop}

% --------------------------------------------------------------------------

\begin{figure}
\rotatebox{270}
{\scalebox{0.32}{\includegraphics{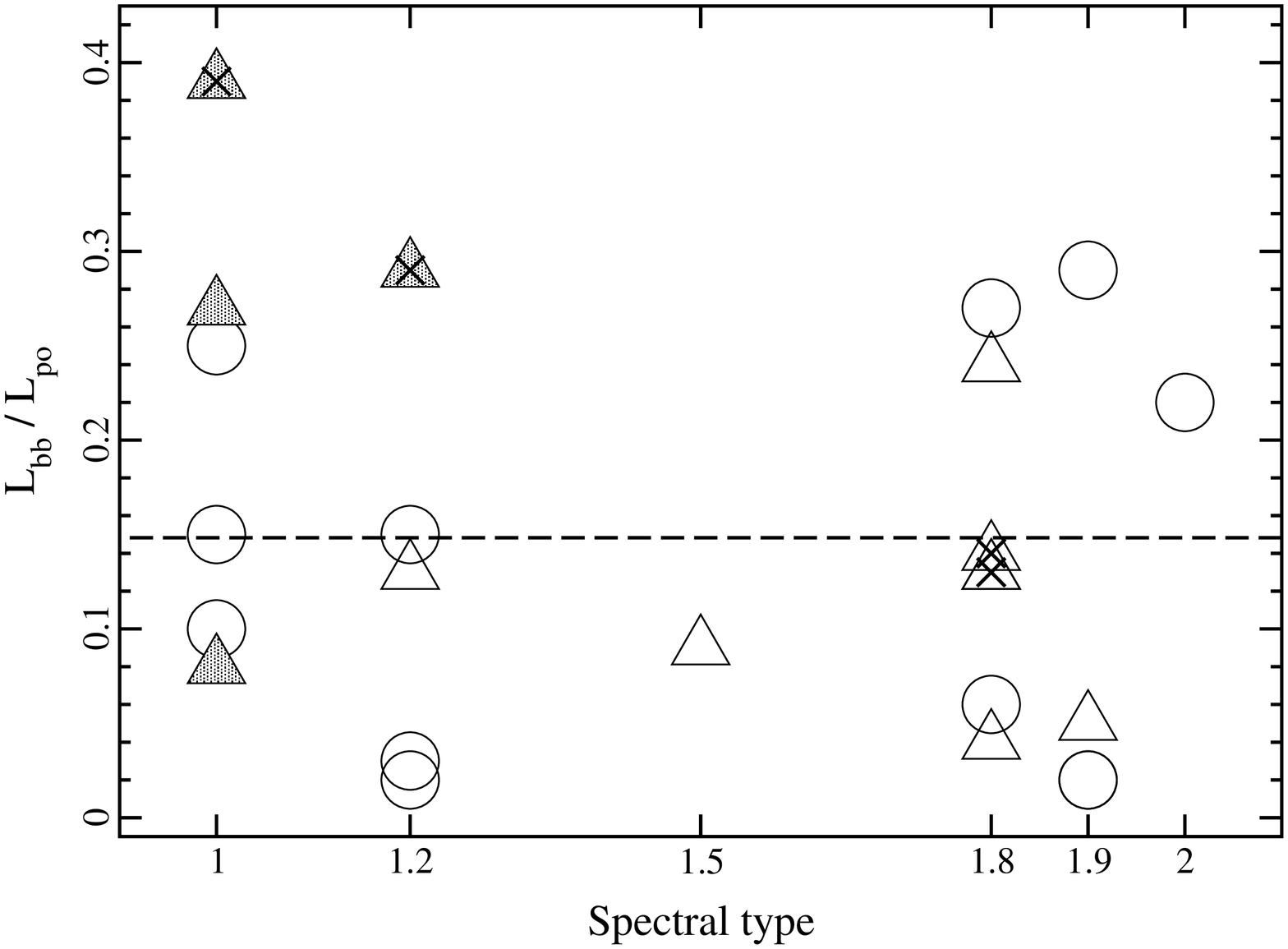}}}
\caption{The luminosity ratio between the blackbody (or Raymond-Smith) and power law
components over the $0.3-10\keV$ band distributed over spectral class.
Triangles mark objects with detected line emission (broad lines
are filled).  AGN fitted with absorption edges are marked with an X.
Objects requiring nothing in addition to a continuum model are shown as circles.
The dashed line marks the average luminosity ratio for the sample.
Some objects have identical ratios; thus overlap in the figure
(see Table~\ref{tab:cont}).
}
\label{fig:samp}
\end{figure}

% --------------------------------------------------------------------------

\subsection{Soft excess emission and neutral intrinsic absorption}
\label{sect:bbc}

It is beyond the scope of this study to make strong claims
about the physical nature of the X-ray continuum, in particular the
nature of the soft excess. We note that in all cases the multi-component
continuum was an improvement to the absorbed power law model, though
not always was it a statistical necessity.   Furthermore, no single
model (i.e. Comptonisation or thermal) stood out as a better fit
to the low-energy spectra in the sample.  Sometimes the
thermal plus power law model was better than the broken power law,
other times the opposite was true.

Notably, there does not appear to be a
clear correlation between the strength of the soft excess and the spectral
classification of AGN in this sample (Figure~\ref{fig:samp}).
Indeed, weak and strong soft excesses seem to be distributed amongst
type 1 and type 2 
(we will include type 1.8 and 1.9 in this group for discussion purposes) objects, 
with the low luminosity type 2 AGN (HCG~4a, NGC~5005, NGC~7331) possessing 
particularly strong soft excesses.
Moreover, there does not appear to be a difference in the average
blackbody temperature between Seyfert 1s and Seyfert 2s 
(excluding the two LINERs), although a significant discrepancy
was recently reported by Mateos \et (2005) for a sample of objects in the 
Lockman Hole.
The average type 1 and type 2 blackbody temperature in our sample is
$kT = 127\pm26$ and $140\pm47\eV$, respectively (the uncertainties in the
average value is the standard deviation in each subclass of objects).

The main difference between the two multi-component models was in the
treatment of the intrinsic cold absorption (ICA).  Although in most cases
the ICA was poorly constrained and only upper limits
could be calculated, there was a tendency for the broken power law 
model to require a greater amount 
of ICA.  This was obviously a means for the model
to accommodate for curvature in the low-energy spectrum of most sources
which was satisfied by the natural curvature of the blackbody function.
However, the differences were never extreme, and the additional
ICA required by the broken power law fits was typically 
$< 10\times$ that measured with the blackbody model. 
We note that there are known uncertainties in the pn calibration in
{\tt SAS v6.1.0} which will result in underestimating the absorption
by a few times $10^{19} \pscm$ (Kirsch 2005).  While this implies that there is
a degree of uncertainty in the values we have reported, the uncertainty is
sufficiently low that we can rule out high levels of absorption.
Indeed none of the objects in this sample required intrinsic absorption
$\gs 10^{21} \pscm$, regardless of the adopted continuum model or calibration
uncertainties.

% --------------------------------------------------------------------------

\subsection{Ionised absorption}
\label{sect:wa}

% --------------------------------------------------------------------------

\begin{figure}
\rotatebox{270}
{\scalebox{0.32}{\includegraphics{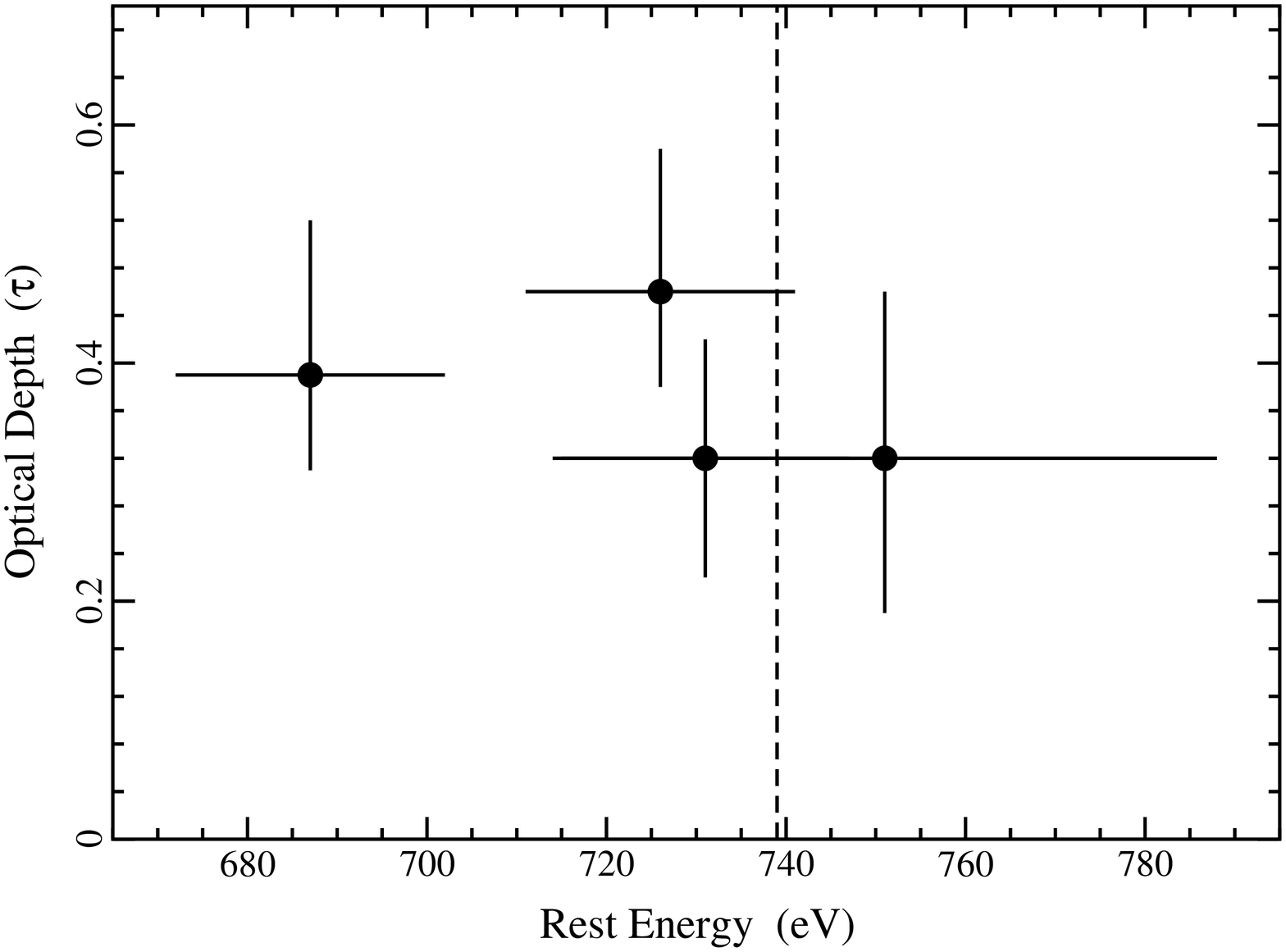}}}
\caption{The optical depths and rest frame energies of the absorption edges
detected in the sample.  The vertical dashed line corresponds to the energy
of the O~\textsc{vii} edge ($739\eV$).
}
\label{fig:edge}
\end{figure}

% --------------------------------------------------------------------------

An absorption edge was a significant addition to the spectra of four AGN in our
sample (Table~\ref{tab:adds}).
With the exception of ESO~602--~G~031, the measured edge energies were all consistent
with absorption from O~\textsc{vii} (Figure~\ref{fig:edge}).
The only distinguishing characteristic in this sub-sample was that
the average luminosity ratio between the blackbody and power law components
is higher than it is for the entire sample (Figure~\ref{fig:samp}).
It is not reasonable to claim that this is a significant result given the small
size of the sample, but it may be worth examining in studies with  
more objects.   

In the case of ESO~602--~G~031, the edge was
slightly redshifted with respect to O~\textsc{vii}. 
McKernan \et (2005) demonstrated that in some objects, redshifted absorption
edges could be associated with absorption by hot local gas since the velocities
measured in the absorber correspond to the recessional velocities of the
AGN.  Assuming O~\textsc{vii} absorption, the measured velocity of the ionised 
absorber in ESO~602--~G~031
is $cz = 21095^{+6490}_{-6085}\kmps$, whereas the recessional velocity of
the AGN is about $10200\kmps$.  This implies that the detected feature in
ESO~602--~G~031 is not local.  However, given the complexity of warm absorbers, the
blending of features, and the moderate resolution of the EPIC cameras 
it is not normally possible to unequivocally identify features, or determine
velocity shifts.

\subsection{Fe\ka\ emission}

\subsubsection{Narrow Fe\ka\ emission}

In one-half of the sample (11 objects from 12 observations) an
emission line was detected in the high-energy spectrum.
In seven objects (ESO~602--~G~031, Mrk~728, UGC~12139,
ESO~244--~G~017, UGC~3973 (first epoch), ESO~15--~IG~011 (second epoch),
CGCG~039--167) the line was consistent with narrow,
unresolved emission from neutral iron.  HCG~4a also showed indications
of a narrow, unresolved emission line, but at 
$E = 6.93^{+0.07}_{-0.14}\keV$, which can be attributed to completely
ionised iron.  This in itself is potentially interesting given the rather low
luminosity observed in HCG~4a (see Table~\ref{tab:cont}).
Six of the eight objects displaying narrow emission lines were type 2 AGN,
consistent with the standard unification model that narrow fluorescence
emission in type 2 AGN originates from reflection off the obscuring torus.

\subsubsection{Broad Fe\ka\ emission}
Despite concentrated effort the realisation of relativistically broadened
Fe~\ka\ emission lines in the X-ray spectra of AGN still remains ambiguous with only
a few clear-cut examples (e.g. Fabian \et 2002; Turner \et 2002).
Matters are further complicated
with the recognition that a possible broad emission feature can be manifested
by partial covering of the central engine by a dense, patchy absorber 
(e.g. Tanaka \et 2004).

Four objects in the sample (UGC~3973 (second epoch), Mrk~590, MCG--02-14-009
(see also Porquet 2005),
II~Zw~177) were fitted better if a broad Gaussian profile
was considered rather than an intrinsically narrow profile 
(see Table~\ref{tab:adds}).
All four objects were type 1 AGN, which is in line with unification models
that suggest we have an unobscured view of broad-line forming regions
(e.g. BLR, outer and inner parts of the accretion disc) in Seyfert 1s.

We considered more elaborate models to describe the nature of these broad 
features,
namely line emission from a disc around Schwarzschild black hole 
(Fabian \et 1989) and a Laor profile produced in a 
disc around a Kerr black hole (Laor 1991).  Not surprisingly, given the 
modest data quality, all profiles worked equally well at fitting the
broad emission features.
%In Table~\ref{tab:broad} a comparison is made in the improvement 
%that each line model provided to the continuum fits.

\begin{figure}
\rotatebox{270}
{\scalebox{0.32}{\includegraphics{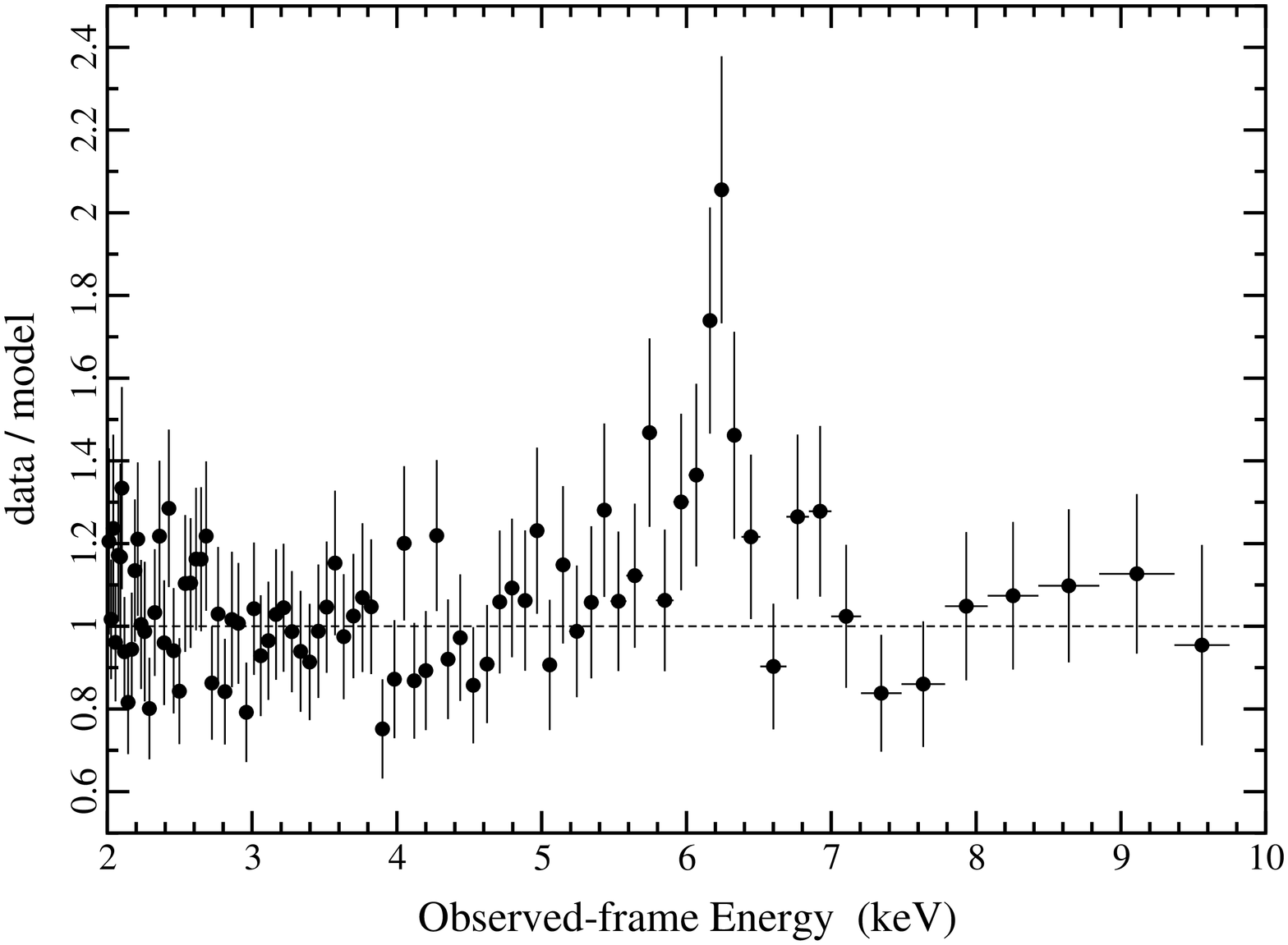}}}
\caption{Residuals (data/model) remaining in the $2-10\keV$
range after fitting the spectrum of Mrk~590 with the
best-fit continuum model discussed in Table~\ref{tab:cont}.
}
\label{fig:mrkln}
\end{figure}

We note that in three objects: MCG--02-14-009, UGC~3973, and Mrk~590
(e.g. Figure~\ref{fig:mrkln}), 
%a simple Gaussian profile satisfactorily fits the excess.  
%The measured energies indicate emission from neutral or slightly ionised
%iron, and 
the full-width at half-maximum of the Gaussian profiles indicated velocities
on the order of $\approx 10000\kmps$.
For all three of these AGN, the widths of the iron lines
are only about a factor of $2-4$ greater then the velocities measured in the
H$\beta$ emission line from the broad line region.  This suggests that, at least in these
cases, the iron emission could be originating from the outer regions of the
accretion disc, where Doppler effects dominate relativistic 
effects.
% --------------------------------------------------------------------------

% --------------------------------------------------------------------------

The profile in the NLS1, II~Zw~177, appeared tantalising.
%ambiguous, and none of the line models appears superior to the others.
The feature was very broad, with $\sigma \approx 1\keV$, and the best-fit
line energy was redshifted from $6.4\keV$,
though not formally inconsistent with neutral iron (see Figure~\ref{fig:line}).

We reiterate that given the available statistics, absorption models
(such as partial covering), which would introduce curvature in the
high-energy spectrum, cannot be formally excluded.  
%However, deeper observations
%of II~Zw~177 would certainly be fruitful to examine effects close to the central
%black hole.

\subsubsection{Unidentified narrow features}
\label{sect:unf}

Recently, a number of emission and absorption features 
have been detected in some AGN, which
appear at energies red- or blue-shifted with respect to Fe~\ka\
(e.g. Matt \et 2005; Turner \et 2004, 2002; Guainazzi 2003; Yaqoob \et 2003).
Proposed explanations include rotating hot spots on the accretion disc
(Iwasawa \et 2004 and references within) and inflowing or outflowing
material (Turner \et 2004) as may be expected from an aborted jet 
(Ghisellini \et 2004).

Based on our definition of a detection, the combined
EPIC spectra did not show any such features.  However, there 
were some potentially interesting features which appeared significant
in the higher signal-to-noise pn spectrum.  
Two AGN which stand out are
ESO~113--~G~010, with a possible narrow line at $\sim5.4\keV$;
and UGC~3973, with a potential line at $E = 7.99\pm0.06\keV$.
Both spectra are presented in detail in other works
(Porquet \et 2004, for ESO~113--~G~010; Gallo \et 2005
for UGC~3973) and we will not discuss them further here.

\subsection{Unabsorbed type 2 AGN}
\label{sect:ut2}
Much work has been exerted on the study of optically defined type 2 AGN
which exhibit X-ray behaviour more typical of type 1 AGN (e.g. unabsorbed
spectra, extreme variability, soft-excess emission; e.g. Risaliti \et 2005;
Barcons \et 2003; Risaliti 2002; PB02; Boller \et 2002).  
Such objects
remain a challenge to describe in terms of the standard orientation-based
unification model. 
In this study, we have identified eleven objects, which are optically defined as 
Seyfert 1.8--2, but possess type 1 AGN X-ray characteristics.

\subsubsection{Pure type 2 AGN}
It has been suggested that there exists a type 2 AGN
dichotomy, in which we have the usual obscured type 1 AGN, along with
``pure'' type 2 AGN (e.g. Tran 2001).  
Tran suggests that in pure Seyfert 2s,
a Seyfert 1 nucleus is overwhelmed by a strong starbust component rendering
the BLR weak or even absent.  
On the other hand, Lumsden \et (2004) dismiss this proposed division,
attributing it largely to data quality.

Nicastro \et (2003) find that pure type 2 AGN candidates 
are very low accretion rate (and low-luminosity) systems, in which the 
BLR does not form (Nicastro 2000).  The low-luminosity aspect was noticed
by PB02, who found a large fraction of 
unabsorbed type 2 objects when low-luminosity AGN 
(LLAGN; $L_{2-10\keV} \approx (0.04-5)\times 10^{41}\ergps$, e.g. 
Terashima \et 2002) were considered.

With the exception of HCG~4a and the two LINERs, the unabsorbed Seyfert 2s in
our sample have $2-10\keV$ luminosities on the order of $10^{42}\ergps$.
This is rather typical for Seyfert activity,
so it is not conclusive that the objects 
presented here are ``pure'' type 2 AGN candidates.  

\subsubsection{High dust-to-gas ratio}
The intrinsic column densities estimated for the type 2 AGN in our sample are
on the order of $10^{20}\pscm$.  Following Gorenstein (1975), this translates
into an extinction of $A_v \approx 0.045$, which is insufficient to redden the
BLR in Seyfert 1.8 and 1.9s ($2 \ls A_v \ls 5$, Goodrich \et 1994).  
To satisfy this discrepancy the galaxies in our sample would require
a gas-to-dust ratio up to 100 times that observed in our Galaxy, which could
occur if the AGN is viewed through some dusty and patchy environment within 
the host galaxy (e.g. dust lanes, bars, and star-forming regions).
A similarly large value was found for the sample analysed by 
PB02, who further suggested that the X-ray absorption
was consistent with arising in the narrow line region, the consequence 
being that a torus was not required.

\subsubsection{Variability}
We note that none of the X-ray observations were contemporaneous with the
optical classification of these objects.  As optical and X-ray spectra
are known to vary with time, there is always a level of uncertainty involved.
An extreme example is the X-ray observations of NGC 1365 (Risaliti \et 2005), in
which the spectrum went from being Compton thick to Compton thin, 
and back to thick, in just 6 weeks.

Risaliti \et proposed a model where the variations are attributed to 
line-of-sight effects, in which the AGN is viewed through a stratified
absorber (see their fig. 2).  
In the case of NGC~1365, the AGN is viewed on the edge where the medium changes
from Compton thick to thin.  Subtle changes in the composition in this mixing
region can produce the observed effects.

A similar model can be envisaged for unabsorbed Seyfert 2s, in general.
Frequent X-ray exposures, and simultaneous optical observations of a few of
the objects in this sample could be enlightening.

\subsubsection{How many unabsorbed Seyfert 2s are there?}
A bright and soft survey as presented here will naturally be biased toward
unabsorbed objected.  Therefore estimating the number of AGN expected to
have similar properties is hardly accurate.  Indeed, $all$ of the type 2 AGN
in our sample showed low levels of intrinsic absorption.

From the deep \xmm\ observations of the Lockman Hole, Mateos \et (2005)
reported that 5 out of 28 optically defined type 2 AGN exhibited low levels of
absorption.  The ratio ($\sim 18\%$) is comparable to the estimate of
$10-30\%$ suggested by PB02.

% --------------------------------------------------------------------------

\subsection{X-ray emission from NGC~5005 and NGC~7331}
\label{sect:ott}

The low-energy spectrum of the two LINERs, NGC~5005 and NGC~7331,
demonstrated more complexity than seen in the other objects of the sample.
When fitted with a simple power law the data showed excess emission concentrated
in the $0.6-1\keV$ range
(Figure~\ref{fig:po}) similar to that observed in the
Seyfert~2, NGC~5643 (Guainazzi \et 2004).

A highly absorbed power law or blackbody continuum, in addition to emission
and absorption lines could improve the fit, but the simplest and best fits were
found when modelling the soft excess with emission from an optically thin
thermal plasma.
The best-fit parameters for NGC~5005 were comparable to those obtained with
earlier \asca\ data (Terashima \et 2002).

The nature of the soft excess in NGC~5005 and NGC~7331 may be unusual in terms
of ``normal'' AGN behaviour, but not uncommon in terms of LINER activity.
The spectral properties of both objects are consistent with the known X-ray
class properties of LINERs (Terashima \et 2002).

The proximity of the LINERs makes it possible to recognise that both objects are
extended in the \xmm\ images.  It has been previously determined that a significant
fraction of the soft emission in NGC~5005 (Rush \& Malkan 1996) and
NGC~7331 (Tyler \et 2004) originates from an extended component on host-galaxy
scales.  In principle, the hard power law could also be due to the contribution
of galactic binaries in the host galaxy.  In fact, Swartz \et (2004) identified three ultraluminous
X-ray (ULX) sources associated with NGC~7331, which could be contributing part of
the high-energy emission.  However, from a \chandra\ image of NGC~7331,
Tyler \et (2004) determined that the peak X-ray emission was coming from the
galactic nucleus and not the vicinity of the molecular ring or host galaxy.  Of course,
this is not proof that the hard X-rays are coming from the Seyfert 2 nucleus,
but at least for NGC~7331 we consider it the most likely scenario.

% --------------------------------------------------------------------------

\section{Conclusions}
\label{sect:conc}

We have presented a snap-shot survey of twenty-one bright, \rosat\
selected AGN observed in the $0.3-10\keV$ band with \xmm.
The main results of the analysis are as follow:

\begin{itemize}

\item
All of the sources, including eleven type 1.8--2 objects,
showed low levels of intrinsic absorption ($\ls 10^{21}\pscm$).
Not surprisingly, a bright soft-X-ray sample such as this may be biased toward
low-absorption systems.  This is beneficial in identifying 
unabsorbed Seyfert 2s, which do not clearly fit into the orientation-based 
unification models.

\item
There does not appear to be a clear correlation between the strength and
shape (i.e. temperature) of the soft
excess, and the spectral classification of AGN in this sample. 

\item
The two LINER type objects in the sample displayed extended X-ray emission
in their \xmm\ image and a soft excess above a power law, which was best
described with optically thin thermal emission.

\item
Low-energy absorption edges are detected in four objects.  In three cases
the best fit edge energies are consistent with O~\textsc{vii} absorption.
In one case, the best-fit energy is redshifted with respect to O~\textsc{vii}.

\item
Four objects showed indications of broad Fe~\ka\ emission, 
all of which were type 1 AGN. 
The lines in three of the objects could be explained by emission
from the outer regions
of an accretion disc as they display velocity widths greater than those
seen in the optical BLR, but not at relativistic levels expected
closer to the black hole.
The feature in the NLS1, II~Zw~177, was very broad and an origin close to the
central black hole, or partial covering,  should be considered.

\end{itemize}

Of primary interest is the detection of a large fraction of unabsorbed 
type 2 AGN.
We have demonstrated that a bright RASS selected sample such as this one can 
be useful in selecting unabsorbed type 2 AGN candidates.
Cross-correlation of the RASS or the \rosat\ Bright Survey
(Schwope \et 2000) with large optically selected samples of 
Seyfert 2s, such as those available with the SDSS, can potentially reveal
hundreds of candidates.  

%The sample includes several individual objects which certainly warrant deeper
%follow-up observations.  We also present several new unabsorbed type 2 AGN
%which should provide a small database for fruitful study of the type 2 AGN
%dichotomy.

% --------------------------------------------------------------------------

% --------------------------------------------------------------------------

\section*{Acknowledgements}

Based on observations obtained with \xmm, an ESA science mission with
instruments and contributions directly funded by ESA Member States and
the USA (NASA).  Many thanks to Karsten Bischoff for providing us with
a copy of his PhD dissertation, and to G\"unther Hasinger for useful 
discussion.  Much appreciation to the anonymous referee for providing
constructive comments, which lead to improvement of the manuscript.

%\pagebreak

\bsp
\label{lastpage}

\begin{thebibliography}{}

 
\bibitem{} Antonucci R. R. J., Miller J. S., 1985, ApJ, 297, 621

\bibitem{2} Arnaud K., 1996, in: {\it Astronomical Data Analysis Software and Systems}, Jacoby G.,  Barnes J., eds, ASP Conf. Series Vol. 101, p17 

\bibitem{} Awaki H., Koyama K., Inoue H., Halpern J., 1991, PASJ, 43, 195

%\bibitem{} Balestra I., Bianchi S., Matt G., 2004, A\&A, 415, 437 

\bibitem{} Barcons X., Carrera F., Ceballos M., 2003, MNRAS, 339, 757

\bibitem{} Bischoff K., 2004, PhD thesis, Univ. G\"ottingen

\bibitem{} Boller Th. \et, 2003, A\&A, 397, 557 

%\bibitem{} Dewangan G., Griffiths R., Schurch N., 2003, ApJ, 592, 52

\bibitem{} Dickey J. M., Lockman F. J., 1990, ARA\&A, 28, 215

\bibitem{} Ehle M. \et 2004, {\em XMM-Newton} Users' Handbook, iss 2.2

\bibitem{} Fabian A. C., Rees M. J., Stella L., White N. E., 1989, MNRAS, 238, 729

\bibitem{} Fabian A. C. \et, 2002, MNRAS, 335, 1

\bibitem{} Freyberg M. J. et al., 2004, in Flanagan K., Siegmund O., eds,
{\it X-Ray and Gamma-Ray Instrumentation for Astronomy XIII},
Proceedings of the SPIE, 5165, 112

\bibitem{} Gallo L., Fabian A., Boller Th., Pietsch W., 2005, MNRAS, 363, 64

\bibitem{} Ghisellini G., Haardt F., Matt G., 2004, A\&A, 413, 535

\bibitem{} Goodrich R., Veilleux S., Hill G., 1994, ApJ, 422, 521

\bibitem{} Gorenstein P., 1975, ApJ, 198, 95

\bibitem{} Guainazzi M., 2003, A\&A, 401, 903

\bibitem{} Guainazzi M., Rodriguez-Pascual P., Fabian A., Iwasawa K., Matt G., 2004,
MNRAS, 355, 297

\bibitem{} Grupe D., Thomas H.-C., Beuermann K., 2001, A\&A, 367, 470

\bibitem{} Grupe D., Wills B., Leighly K., Meusinger H., 2004, AJ, 127, 156

\bibitem{} Ho L. C., Filippenko A. V., Sargent W. L. W., 1997, ApJS, 112, 315

\bibitem{} Iwasawa K., Miniutti G., Fabian A., 2004, MNRAS, 355, 1073

\bibitem{} Jansen F. \et 2001, A\&A, 365, L1 

\bibitem{}  Kaastra J.S., 1992, {\it An X-Ray Spectral Code for Optically Thin Plasmas}
(Internal SRON-Leiden Report, updated version 2.0)

\bibitem{} Kewley L. J., Heisler C. A., Dopita M. A., Lumsden S., 2001, 
ApJS, 132, 37

\bibitem{} Kirsch M., 2005, \xmm\ Calibration Documents
(CAL-TN-0018-2.4)

\bibitem{} Kuraszkiewicz J., Green P., Crenshaw D., Dunn J., Forster K., 
Vestergaard M., Aldcroft T., 2004, ApJS, 150, 165

\bibitem{} Laor A., 1991, ApJ, 376, 90

\bibitem{} Lumsden S., Alexander D., Hough J., 2004, MNRAS, 348, 1451

\bibitem{} Mewe R., Gronenschild E.H.B.M., van den Oord G.H.J., 1985,
A\&AS, 62, 197

\bibitem{} Mewe R., Lemen J.R., van den Oord G.H.J., 1986, A\&AS, 65, 511

\bibitem{} Nicastro F. 2000, ApJ, 530, 65

\bibitem{} Nicastro F., Martocchia A., Matt G., 2003, ApJ, 589, 13

\bibitem{} Matt G., Porquet D., Bianchi S., Falocco S., Maiolino R., 
Reeves J., Zappacosta L., 2005, Accepted by A\&A, (astro-ph/0502323)

\bibitem{} Matt G., Guainazzi M., Maiolino R., 2003, MNRAS, 342, 422

\bibitem{} Mattson B., Weaver K., 2004, ApJ, 601, 771 

\bibitem{} McKernan B., Yaqoob T., Reynolds C., 2004, ApJ, 617, 232

\bibitem{} Moran E., Kay L., Davis M., Filippenko A., Barth A., 2001, 
ApJ, 556, 75

\bibitem{} Osterbrock D. E., Martel A., 1993, ApJ, 414, 552

\bibitem{} Osterbrock D. E., Shuder J. M., 1982, ApJS, 49, 149

\bibitem{} Panessa F., Bassani L., 2002, A\&A, 394, 435 (PB02) 

\bibitem{} Pappa A., Georgantopoulos I., Stewart G., Zezas A., 2001, 
MNRAS, 326, 995

\bibitem{} Paturel G., Fouque P., Bottinelli L., Gouguenheim L., 1989, A\&AS, 80, 299

\bibitem{} Pietsch W., Bischoff K., Boller Th., D\"obereiner S., Kollatschny W., Zimmermann H.-U.,
1998, A\&A, 333, 48

\bibitem{} Porquet D., Reeves J, Uttley P., Turner T. J., 2004, A\&A, 427, 101

\bibitem{} Porquet D., 2005, submitted to A\&A

\bibitem{} Protassov R., van Dyk D., Connors A., Kashyap V., Siemiginowska A., 2002, ApJ, 571, 545

\bibitem{} Raymond J. C., Smith B. W., 1977, ApJS, 35, 419 

\bibitem{}
Risaliti G., 2002, A\&A, 386, 379

\bibitem{}
Risaliti G., Elvis M., Fabbiano G., Baldi A., Zezas A., 2005, ApJ, 623, 93

\bibitem{} Ross R. R., Fabian A. C., Brandt W. N., 1996, MNRAS, 278, 1082

\bibitem{} Rush B., Malkan M., 1996, ApJ, 456, 466  

\bibitem{} Schwope A. \et, 2000, AN, 321, 1

\bibitem{} Str\"{u}der L. \et 2001, A\&A, 365, L18 

\bibitem{} Swartz D., Ghosh K., Tennant A., Wu K., 2004, ApJS, 154, 519

\bibitem{} Tanaka Y.; Boller Th., Gallo L., Keil R., Ueda Y., 2004, PASJ, 56, 9

\bibitem{} Terashima Y., Iyomoto N., Ho L., Ptak A., 2002, ApJS, 139, 1 

\bibitem{} Tran H., 2001, ApJ, 554, 19 

\bibitem{166}
Turner M. J. \et, 2001, A\&A, 365, 27

\bibitem{} Turner T. J. \et 2002, ApJ, 574, L123

\bibitem{} Turner T. J., Kraemer S. B., Reeves J. N., 2004, ApJ, 603, 62

\bibitem{} Tyler K., Quillen A., LaPage A., Rieke G., 2004, ApJ, 610, 213

\bibitem{} Veron-Cetty M., Veron P., 2003, A\&A, 412, 399 

\bibitem{} Voges W. \et, 1999, A\&A, 349, 389 

\bibitem{} Wandel A., Peterson B., Malkan M., 1999, ApJ, 526, 579

%\bibitem{}
%Vaughan S., Reeves J., Warwick R., Edelson R., 1999, MNRAS, 309, 113

\bibitem{}
Yaqoob T., George I. M., Kallman T. R., Padmanabhan U., Weaver K. A.,
Turner T. J., 2003, ApJ, 596, 85

\bibitem{} Zimmermann H.-U., Boller Th., D\"obereiner S., Pietsch W., 2001,
A\&A, 378, 30

\end{thebibliography}
\end{document}